\documentstyle[prl,aps,twocolumn,eqsecnum]{revtex}

\preprint{Submitted to {\it Physical Review A}}

\draft

\author{M. A. Nielsen\thanks{%
Electronic address: mnielsen@tangelo.phys.unm.edu}%
and Carlton M.~Caves}

\title{Reversible quantum operations and their application to teleportation}

\address{Center for Advanced Studies, Department of Physics and Astronomy,\\
 University of New Mexico, Albuquerque, NM 87131-1156}

\date{\today}

\begin{document}

\pagestyle{plain}
\pagenumbering{arabic}

\maketitle

\begin{abstract}
Quantum operations provide a general description of the state changes 
allowed by quantum mechanics.  Simple necessary and sufficient conditions 
for an ideal quantum operation to be reversible by a unitary operation 
are derived in this paper.  These results generalize recent work on 
reversible measurements by Mabuchi and Zoller [Phys.\ Rev.\ Lett.\ 
{\bf 76}, 3108 (1996)].  Quantum teleportation can be understood as a 
special case of the problem of reversing quantum operations.  We 
characterize completely teleportation schemes of the type proposed 
by Bennett {\it et al.}\ [Phys.\ Rev.\ Lett.\ {\bf 70}, 1895 (1993)].
\end{abstract}



\section{Introduction}

Recently Mabuchi and Zoller \cite{Mabuchi96a} have shown how
a measurement on a quantum system can be reversed under appropriate 
conditions. In this paper we derive a simple necessary and sufficient 
condition for an ideal quantum operation to be reversible by a unitary 
operation. Since all quantum measurements can be described by a set of
quantum operations, our result generalizes the scheme of Mabuchi and
Zoller.  Moreover, it shows how the reversibility of a measurement is 
connected to the information gained from that measurement. 

This paper also examines the teleportation of quantum states, first 
discussed by Bennett {\it et al.}\ \cite{Bennett93a}. We show that 
teleportation can be recast as the problem of reversing a set of 
quantum operations.  The necessary and sufficient condition for 
unitarily reversing an ideal quantum operation is then applied to 
give a complete characterization of teleportation schemes of the type 
proposed by Bennett {\it et al.}

The structure of the paper is as follows. In Sec.~\ref{sect: qop},
after reviewing the formalism of quantum operations, we define ideal 
quantum operations and unitarily reversible quantum operations. In 
Sec.~\ref{sect: qmeas} we review how quantum measurements can be 
described in terms of a set of quantum operations and show that the 
scheme of Mabuchi and Zoller is an example of a unitarily reversible 
ideal quantum measurement.  Section \ref{sect: characterize} contains 
the statement and proof of a general characterization of unitarily 
reversible ideal quantum operations; this characterization generalizes 
the results of Mabuchi and Zoller.  Section~\ref{sect: teleport}, 
after reviewing the teleportation scheme described by Bennett 
{\it et al.}\ \cite{Bennett93a}, formulates the general problem of 
teleportation and shows how it can be understood as a special case of 
the problem of reversing a set of quantum operations. Finally, in 
Sec.~\ref{sect: Bennett} we use the condition for unitarily reversing 
an ideal measurement to give a complete characterization of teleportation 
schemes of the type proposed in \cite{Bennett93a}. A concluding 
Sec.~\ref{sect: conc} summarizes our results.

\section{Quantum Operations}
\label{sect: qop}

A simple example of a state change in quantum mechanics is
the unitary evolution experienced by a closed quantum system. 
The final state of the system is related to the initial
state by a unitary transformation $U$,
\begin{eqnarray} \label{eqtn: unitary operation}
\rho \rightarrow {\cal E}(\rho) = U \rho U^{\dagger}\;. \end{eqnarray}
Unitary evolution is not the most general type of state change possible 
in quantum mechanics. Other state changes, not describable by unitary 
transformations, arise when a quantum system is coupled to an environment 
or when a measurement is performed on the system.

How does one describe the most general possible state change in 
quantum mechanics? The answer to this question is provided by the 
formalism of ``quantum operations.''  This formalism is described in 
detail by Kraus \cite{Kraus83a} and is given a short, but quite 
informative review by Schumacher in an Appendix \cite{Schumacher96a}. 
In this formalism there is an {\em input state\/} and an {\em output 
state}, which are connected by a map
\begin{eqnarray}
\rho \rightarrow 
\frac{{\cal E}(\rho)}{\mbox{tr}\bigl({\cal E}(\rho)\bigr)}\;.
\end{eqnarray}
This map is determined by a {\em quantum operation\/} ${\cal E}$, 
a linear, trace-decreasing map that preserves positivity.  The trace 
in the denominator is included in order to preserve the trace condition,
$\mbox{tr}(\rho) = 1$.

The most general form for ${\cal E}$ that is physically reasonable 
(in addition to being linear and trace-decreasing and preserving
positivity, a physically reasonable ${\cal E}$ must satisfy an 
additional property called complete positivity),
can be shown to be \cite{Kraus83a}
\begin{eqnarray} \label{eqtn: gen op}
{\cal E}(\rho) = \sum_j A_j \rho A_j^{\dagger}\;. \end{eqnarray}
The system operators $A_j$, which must satisfy $\sum_j A_j^\dagger A_j\le I$, 
completely specify the quantum operation.  In the particular case of a 
unitary transformation, there is only one term in the sum, $A_1 = U$, 
leaving us with the transformation~(\ref{eqtn: unitary operation}).

We now come to the two definitions that delineate the class of reversal
problems considered in this paper.  We say that a quantum operation 
${\cal E}$ is {\em ideal\/} if it can be written in the form
\begin{eqnarray}
{\cal E}(\rho) = A \rho A^{\dagger} \end{eqnarray}
for some single operator $A$.  The reason for this terminology becomes
more apparent in the next section where we discuss ideal quantum 
measurements.  

When we talk about reversing a quantum operation ${\cal E}$, we generally 
do not mean that ${\cal E}$ can be reversed for all input states, but
rather only that ${\cal E}$ can be reversed for a certain class of input
states, in particular, for all density operators $\rho$ whose support
lies in a subspace $M$ of the total state space $L$.  It makes sense to
talk about reversing ${\cal E}$ on a subspace $M$ only if 
${\cal E}(\rho)\ne0$ for all $\rho$ whose support lies in $M$, and we 
assume this condition henceforth.  We say that a quantum operation 
${\cal E}$ is {\em unitarily reversible\/} on a subspace $M$ if there 
exists a unitary operator $U$, acting on the total state space $L$, such 
that
\begin{eqnarray}
\rho = 
U \frac{{\cal E}(\rho)}{\mbox{tr}\bigl({\cal E}(\rho)\bigr)} U^{\dagger} 
\end{eqnarray}
for all $\rho$ whose support lies in $M$.  

In this paper our main concern is with unitarily reversing ideal quantum 
operations.  Thus we place two restrictions on the class of reversal 
problems that we consider: the restriction to reversing {\em ideal\/} 
operations and the restriction that any reversal be accomplished 
{\em unitarily}.

In principle it is possible to reverse an operation using more general
operations than unitary ones.  The reversing operation must be 
deterministic; it turns out that the most general form for a 
deterministic quantum operation can be obtained by adjoining an 
ancilla system to the system of interest, allowing the system plus 
ancilla to interact unitarily, and then discarding the ancilla.  Such 
a dynamics leads to a state change of the form
\begin{eqnarray} \label{eqtn: detop}
\rho \rightarrow 
\mbox{tr}_A\! \bigl ( V (\rho \otimes \sigma) V^{\dagger} \bigr )
\equiv {\cal R}(\rho)\;,
\end{eqnarray}
where $\mbox{tr}_A$ denotes tracing out the ancilla, $\sigma$ is the 
initial state of the ancilla, and $V$ is the unitary operator for the
joint dynamics of the system and ancilla.  

Such a quantum operation ${\cal R}$, called a {\em deterministic\/} or 
{\em trace-preserving\/} operation, can always \cite{Kraus83a,Schumacher96a} 
be written in the form
\begin{eqnarray}
{\cal R}(\rho) = \sum_j R_j \rho R_j^{\dagger}\;, \end{eqnarray}
where the system operators $R_j$ satisfy the completeness relation
$\sum_j R_j^{\dagger} R_j = I$.  The completeness relation implies
that $\mbox{tr}\bigl({\cal R}(\rho)\bigr)=\mbox{tr}(\rho)=1$, thus 
accounting for the absence of a trace factor to normalize 
Eq.~(\ref{eqtn: detop}).  The deterministic operation ${\cal R}$
reverses ${\cal E}$ on the subspace $M$ if
\begin{eqnarray} \label{eqtn: gen rev}
\rho = 
{\cal R}\!\left(
\frac{{\cal E}(\rho)}{\mbox{tr}\bigl({\cal E}(\rho)\bigr)}\right)
\end{eqnarray}
for all $\rho$ whose support lies in $M$.

Even though we restrict the reversal problems considered here to 
unitarily reversing ideal operations, the restricted problem is still
of considerable interest.  We show that both the results of Mabuchi 
and Zoller \cite{Mabuchi96a} and the teleportation scheme of Bennett 
{\it et al.}\ \cite{Bennett93a} fit within this framework. 

\section{Quantum Measurements}
\label{sect: qmeas}

Standard textbook treatments describe quantum measurements in terms of
a complete set of orthogonal projection operators for the system being 
measured.  This formalism, however, does not describe many of the actual 
measurements that can be performed on a quantum system. The most general 
type of measurement that can be performed on a quantum system is known as 
a {\em generalized measurement} \cite{Kraus83a,Gardiner91a}.

Generalized measurements can be understood within the framework of 
quantum operations. The most general type of quantum measurement is 
described by a set of system operators $A_{ij}$, labeled by two indices, 
$i$ and $j$, and satisfying the completeness relation
\begin{eqnarray} \label{eqtn: completeness}
\sum_{ij} A_{ij}^{\dagger} A_{ij} = I\;. \end{eqnarray}
The first index, $i$, labels the outcome of the measurement.  If result 
$i$ occurs, then the unnormalized state of the system immediately after 
the measurement is given by
\begin{eqnarray}
{\cal E}_i(\rho) \equiv 
\sum_j A_{ij} \rho A_{ij}^{\dagger}\;. \end{eqnarray}
For each measurement result $i$, a different quantum operation 
${\cal E}_i$ describes the corresponding state change. 

The probability for result $i$ to occur is
\begin{eqnarray}
\mbox{Pr}(i) = \mbox{tr}\bigl({\cal E}_i(\rho)\bigr) =
    \mbox{tr}\!\left(\rho\sum_j A_{ij}^{\dagger}A_{ij}\right). 
\end{eqnarray}
Notice that the measurement probabilities are specified by the positive 
operators
\begin{eqnarray}
E_i\equiv \sum_j A_{ij}^{\dagger} A_{ij}\;. \end{eqnarray}
The normalization condition, $\sum_i \mbox{Pr}(i) = 1$ for all 
density operators, is equivalent to the completeness 
condition~(\ref{eqtn: completeness}), which can be rewritten in terms
of the operators $E_i$ as
\begin{eqnarray} \label{eqtn: completeness2}
\sum_i E_i=I\;. \end{eqnarray} 
A set of positive operators that satisfy the completeness 
relation~(\ref{eqtn: completeness2}) is called a 
{\it positive-operator-valued measure\/} (POVM) \cite{Peres93a}; the
individual operators $E_i$ are called {\it POVM elements}.

We say a measurement is {\em ideal\/} if for each measurement result 
$i$, the corresponding quantum operation ${\cal E}_i$ is ideal; that 
is, there exist operators $A_i$ such that
\begin{eqnarray} \label{ideal state change}
{\cal E}_i(\rho) = A_i \rho A_i^{\dagger}\;. \end{eqnarray}
The probability that result $i$ occurs is given by
\begin{eqnarray}
\mbox{Pr}(i) = \mbox{tr}(\rho A_i^{\dagger}A_i) = 
\mbox{tr}(\rho E_i)\;,\end{eqnarray}
where $E_i=A_i^{\dagger} A_i$ is the POVM element for outcome $i$.  It 
can be shown that ideal measurements correspond in a certain sense to 
doing a perfect readout of the state of the apparatus to which the 
system is coupled.  This is the reason we call such a measurement 
ideal.

For ideal quantum operations the state change (\ref{ideal state change})
corresponding to outcome~$i$ can be described in terms of state vectors 
alone.  It becomes
\begin{eqnarray}
|\psi\rangle \rightarrow \frac{A_i |\psi\rangle}{\sqrt{\langle \psi|
	A_i^{\dagger} A_i |\psi\rangle}}\;, \end{eqnarray}
where the output state vector on the right is only defined up to an 
arbitrary phase factor.  The probability for outcome~$i$ takes the form
\begin{eqnarray}
\mbox{Pr}(i) = 
\langle \psi| A_i^{\dagger} A_i |\psi\rangle = 
\langle \psi| E_i |\psi\rangle 
\;. \end{eqnarray}
This equivalent description in terms of state vectors is often convenient, 
especially to simplify notation; we use it frequently in the following.

We say that a measurement is {\it unitarily reversible\/} on a subspace 
$M$ of the total state space $L$ if for each measurement result $i$, the 
corresponding quantum operation is unitarily reversible; that is, there 
exists a unitary operator $U_i$ such that
\begin{eqnarray} \label{eqtn: urevmeasurement}
U_i \frac{{\cal E}_i(\rho)}{\mbox{tr}\bigl({\cal E}_i(\rho)\bigr)}
U_i^{\dagger} = \rho \end{eqnarray}
for all states $\rho$ whose support lies in $M$.  Outcomes that have zero 
probability on $M$ are irrelevant, because they never occur, and can be 
discarded; recall that for the other outcomes we assume that 
${\cal E}_i(\rho)\ne0$ for all $\rho$ whose support lies in $M$.

If the measurement is ideal, then the quantum operations in 
Eq.~(\ref{eqtn: urevmeasurement}) have the form~(\ref{ideal state change})
involving a single operator $A_i$.  For ideal measurements the equivalent 
definition of a unitarily reversible measurement in terms of state vectors 
is that for all states $|\psi\rangle$ in the subspace $M$,
\begin{eqnarray}
U_i \frac{A_i |\psi\rangle}{\sqrt{\langle \psi|
	A_i^{\dagger} A_i |\psi\rangle}} = |\psi\rangle\;, \end{eqnarray}
where equality here is understood to mean equality up to a phase factor.
Physically, if the initial state lies in the subspace $M$ and result $i$ 
occurs, then applying the unitary operator $U_i$ to the system returns 
it to the state it was in before the measurement.

We could define measurements that are only sometimes reversible by 
requiring that only some of the measurement results have unitarily
reversible quantum operations.  Although we do not deal explicitly with 
sometimes reversible measurements in this paper, the results in 
Sec.~\ref{sect: characterize}, since they are derived for individual 
ideal quantum operations, apply to sometimes reversible measurements.

The scheme proposed by Mabuchi and Zoller \cite{Mabuchi96a} is a particular
type of unitarily reversible ideal measurement, which can be described as 
follows.  Suppose $a$ and $b$ are annihilation operators for two modes 
of the electromagnetic field.  It is possible in principle to perform an
ideal measurement that is described by the following three measurement 
operators:
\begin{eqnarray} 
A_1 & = & \sqrt{\frac{\Delta}{2}}( a + b ) \;, 
\label{a1} \\
A_2 & = & \sqrt{\frac{\Delta}{2}}( a - b ) \;, 
\label{a2} \\
A_3 & = & e^{-ih\Delta}\sqrt{I - \Delta(a^\dagger a+b^\dagger b)} 
\nonumber \\
    & = & I-ih\Delta - {\Delta\over2}(a^\dagger a+b^\dagger b)\;. 
\label{a3} \end{eqnarray}
Here $\Delta$ is an infinitesimal dimensionless time (in the Mabuchi-Zoller 
scheme, $\Delta$ is an infinitesimal time measured in units of a cavity 
damping time), and $h$ is a dimensionless Hamiltonian for the modes $a$
and $b$.  

The measurement described by the operators (\ref{a1})--(\ref{a3}) is 
unitarily reversible on the two-dimensional subspace $M$ spanned by the 
vectors $|2_a 0_b\rangle$ and $|0_a 2_b\rangle$.  Suppose the system is 
initially in an arbitrary state in $M$,
\begin{eqnarray} \label{eqtn: initial state}
|\psi\rangle = \alpha |2_a 0_b \rangle + \beta |0_a 2_b \rangle\;. 
\end{eqnarray}
Since
\begin{eqnarray}
A_1|\psi\rangle&=&\sqrt{\Delta}
\bigl(\alpha |1_a 0_b \rangle + \beta |0_a 1_b \rangle\bigr)\;, \\
A_2|\psi\rangle&=&\sqrt{\Delta}
\bigl(\alpha |1_a 0_b \rangle - \beta |0_a 1_b \rangle\bigr)\;, \\
A_3|\psi\rangle&=&e^{-ih\Delta}\sqrt{1-2\Delta}\,|\psi\rangle\;, \end{eqnarray}
results 1 and 2 each occur with probability $\Delta$, result 3 with 
probability $1-2\Delta$, and the post-measurement states for the 
three results are given by
\begin{eqnarray}
\mbox{Result~1:\quad}&\mbox{}&
\alpha |1_a 0_b \rangle + \beta |0_a 1_b \rangle\;, \\
\mbox{Result~2:\quad}&\mbox{}&
\alpha |1_a 0_b \rangle - \beta |0_a 1_b \rangle\;, \\
\mbox{Result~3:\quad}&\mbox{}&
e^{-ih\Delta}|\psi\rangle\;.\end{eqnarray}
It is easy to see that for each measurement result, the original state 
can be restored by application of an appropriate unitary operation, and 
Mabuchi and Zoller outline a physically plausible process describing how 
this unitary operation can be performed in practice when $h=0$.
Yet why the measurement can be reversed for initial states of the 
form~(\ref{eqtn: initial state}) appears somewhat mysterious in the 
present example.  We now turn to a general result that shows why this 
is the case.


\section{Characterization of Unitarily Reversible Ideal Quantum Operations}
\label{sect: characterize}

In this section we demonstrate that the following conditions are equivalent.

\begin{enumerate}

\item The ideal quantum operation ${\cal E}(\rho) = A \rho A^{\dagger}$
is unitarily reversible on a subspace $M$ of the total state space $L$.

\item The operator $A^\dagger A=E$, when restricted to the subspace $M$,
is a positive multiple of the identity operator on $M$; that is, 
\begin{eqnarray}
P_M A^\dagger AP_M = P_M E P_M =\mu^2 P_M\;, \end{eqnarray}
where $\mu$ is a real constant satisfying $0<\mu\le1$ and $P_M$ is the 
projector onto $M$.

\item The quantity 
$\langle \psi| A^{\dagger} A |\psi\rangle=\langle\psi|E|\psi\rangle$ is a 
positive constant $\mu^2$ for all normalized states $|\psi\rangle$ in $M$, 
where $\mu$ is the real constant of condition 2.  If ${\cal E}$ represents 
a measurement result, this means that the probability of occurrence of 
the result represented by $A$ is the same for all states in $M$.  
Equivalently, $\mbox{tr}(\rho A^{\dagger} A)=\mbox{tr}(\rho E)=\mu^2$ 
for all density operators whose support lies in $M$.

\item The operator $A$ can be written in the form
\begin{eqnarray}
A = \mu V P_M + A P_N\;, \end{eqnarray}
where $V$ is some unitary operator on the whole space $L$, $\mu$ is
the real constant of condition~2, and $P_M$ and $P_N$ are the projectors 
onto the subspaces $M$ and $N$, respectively, where $L = M \oplus N$.
 
\end{enumerate}

\noindent
Note first that conditions 2 and 3 are equivalent, 2 being just a 
restatement of 3 in operator language.  In order to prove the other 
equivalences, we show that $1$ implies $2$ implies $4$ implies $1$.  
Since $4$ implies $1$ can be obtained trivially by using 
$U \equiv V^{\dagger}$ to unitarily reverse the measurement, we only need 
to prove the other two implications.

{\it Condition $1$ implies condition $2$.} For notational convenience 
define $B \equiv P_M A^{\dagger} A P_M$. Considered as an operator on $M$, 
$B$ is Hermitian and satisfies $0<B\le I$; that is, on $M$, $B$ has a 
complete set of orthonormal eigenvectors with eigenvalues in the 
interval $(0,1]$.  Suppose that $|1\rangle$ and $|2\rangle$ are two 
such eigenvectors, with eigenvalues $a_1$ and $a_2$.  Then from condition 
$1$ we have that
\begin{eqnarray}
U A |1\rangle = \sqrt{a_1} |1\rangle \quad\mbox{and}\quad
U A |2\rangle = \sqrt{a_2} |2\rangle\;. \end{eqnarray}
Applying condition~1 to 
$|\psi\rangle \equiv \bigl(|1\rangle + |2\rangle\bigr)/\sqrt{2}$ yields 
\begin{eqnarray}
U A |\psi\rangle = 
\sqrt{\frac{a_1+a_2}{2}}|\psi\rangle =
\frac{\sqrt{a_1 + a_2}}{2} \bigl(
	 |1\rangle + |2\rangle \bigr)\;, \label{contra eqtn 1} \end{eqnarray}
but since $U A$ is linear, we also have that
\begin{eqnarray}
U A |\psi\rangle &=& 
\frac{1}{\sqrt2}\bigl(U A |1\rangle + U A|2\rangle\bigr) \nonumber \\
&=& \frac{1}{\sqrt2}\bigl(\sqrt{a_1}|1\rangle+\sqrt{a_2}|2\rangle\bigr)\;.
	\label{contra eqtn 2} \end{eqnarray}
Comparing (\ref{contra eqtn 1}) and (\ref{contra eqtn 2}) tells us that 
$a_1 = a_2$ and thus that all the eigenvalues of $B$, considered as an
operator on $M$, have the same value $\mu^2$.  It follows that 
$B=\mu^2 P_M$.

{\it Condition $2$ implies condition $4$.} Again define 
\begin{eqnarray} \label{B id}
B \equiv P_M A^{\dagger} A P_M = \mu^2 P_M\;. \end{eqnarray}
{}From $P_M + P_N = I$ we obtain the identity
\begin{eqnarray} \label{trivial id}
A = A P_M + A P_N\;. \end{eqnarray}
The polar decomposition theorem [see Eq.~(3.74) of \cite{Peres93a}] 
implies that there exists a unitary operator $V$ on $L$ such that
\begin{eqnarray} \label{eqtn: polar}
A P_M = V \sqrt{P_M A^{\dagger} A P_M}=V \sqrt B\;, \end{eqnarray}
and from (\ref{B id}), (\ref{trivial id}), and (\ref{eqtn: polar}),
we see that
\begin{eqnarray}
A = \mu V P_M + A P_N\;. \end{eqnarray}
This completes the proof.

\vspace{2mm}

It is easy to check that the scheme of Mabuchi and Zoller is an instance
of the general result.  As we have already noted, for all states in the
subspace $M$, results 1 and 2 each occur with probability $\Delta$ and 
result 3 with probability $1-2\Delta$.  The equivalence of conditions 1 
and 3 then implies that the state change for each outcome can be reversed 
by some unitary operator.

Condition~4 makes clear formally why an ideal operation described by $A$ 
can be unitarily reversed on $M$: when acting on states in $M$, $A$ acts 
like the unitary operator $V$, except for rescaling by the real constant 
$\mu$, which accounts for the probability of obtaining the result 
corresponding to $A$.

The physical meaning of condition~3 is clear for the set of ideal
quantum operations that describe an ideal measurement: an ideal 
measurement is reversible if and only if no information about the identity 
of the prior state is obtained from the measurement; more precisely, 
no inference about the prior state in $M$ can be made, since each state 
is equally likely, given any result $i$.  The necessity of condition~3 
for reversing a measurement is obvious: if one could obtain information 
about the prior state and then restore the prior state, then by repeating 
the measurement and restoration many times, one could obtain enough
information to distinguish nonorthogonal states reliably.  The necessity 
of condition~3 for reversing a {\it single\/} quantum operation, though 
plausible on the same grounds, is not obvious.  Moreover, the important 
feature of our result is not the necessity, but rather the sufficiency 
of condition~3 for unitarily reversing an ideal quantum operation.

In view of these remarks it should not be surprising that we can extend
the result that condition~1 implies conditions~2 and 3 to apply to 
deterministic reversal of a general quantum operation, not just unitary 
reversal of an ideal quantum operation.  Besides being of interest in 
its own right, this extension is used later to show that a necessary 
condition for teleportation is that the teleportation process obtain 
no information about the state to be teleported. 

Suppose that $\cal E$ is a general quantum operation, specified by 
operators $A_j$ as in Eq.~(\ref{eqtn: gen op}), and that ${\cal E}$ 
can be reversed by the deterministic operation $\cal R$ on a subspace 
$M$ of the total state space $L$; that is, Eq.~(\ref{eqtn: gen rev}) 
holds for all density operators $\rho$ whose support lies in $M$.
The operator
\begin{eqnarray}
B \equiv P_M\!\left(\sum_j A_j^{\dagger} A_j \right) P_M\;, \end{eqnarray}
considered as an operator on $M$, is Hermitian and satisfies
$0 < B \leq I$, so $B$ has a complete set of orthonormal eigenvectors
on $M$, with eigenvalues in the interval $(0,1]$.  Notice that 
$\mbox{tr}({\cal E}(\rho)) = \mbox{tr}(\rho B)$ for all density operators
$\rho$ whose support lies in $M$. 

Suppose that $|1\rangle$ and $|2\rangle$ are any two orthonormal eigenvectors
of $B$ in $M$, with eigenvalues $a_1$ and $a_2$.  Define 
$\rho_1 \equiv |1\rangle\langle 1|, \rho_2 \equiv |2\rangle \langle 2|$,
and $\rho' \equiv \frac 12 (\rho_1 + \rho_2)$; notice that the support of
each of these density operators lies in $M$.  Condition 
(\ref{eqtn: gen rev}) implies that
\begin{eqnarray}
{\cal R} \bigl( {\cal E}(\rho_1) \bigr) & = & a_1 \rho_1\;, \\
{\cal R} \bigl( {\cal E}(\rho_2) \bigr) & = & a_2 \rho_2\;, \\
\label{eqtn: nec inter}
{\cal R} \bigl( {\cal E}(\rho ') \bigr) & = & 
	\frac 12 (a_1+a_2) \frac 12 (\rho_1 + \rho_2)\;,
\end{eqnarray}
but from the linearity of ${\cal R}$ and ${\cal E}$, we also have that
\begin{eqnarray} \label{eqtn: just}
{\cal R} \bigl( {\cal E}(\rho ') \bigr) & = & \frac 12 \bigl[
	{\cal R} \bigl( {\cal E}(\rho_1) \bigr) + 
	{\cal R} \bigl( {\cal E} (\rho_2) \bigr) \bigr] \nonumber \\
& = & \frac 12 (a_1 \rho_1 + a_2 \rho_2).
\end{eqnarray}
Comparing Eqs.~(\ref{eqtn: nec inter}) and (\ref{eqtn: just}), we see
that $a_1 = a_2$.  Thus all the eigenvalues of $B$, considered as
an operator on $M$, have the same value $\mu^2$, where $0<\mu\le1$;
that is, 
\begin{eqnarray}
B= P_M\!\left(\sum_j A_j^{\dagger} A_j \right) P_M=
\mu^2 P_M\;. \end{eqnarray} 
Equivalently, we see that $\mbox{tr}\bigl(\cal E(\rho)\bigr)$ is a 
positive constant, $\mu^2$, for all density operators $\rho$ whose 
support lies in $M$.  If $\cal E$ represents a measurement result, 
this means that the probability of the result represented by $\cal E$ 
is the same for all states whose support lies in $M$.

Thus we see that a necessary condition for reversing a general quantum 
measurement is that no information about the identity of the prior
state be obtained from the measurement.  It is clear that this is not
also a sufficient condition, since it is easy to construct nonideal quantum
operations for which $\mbox{tr}\bigl({\cal E}(\rho)\bigr)$ is the same
for all states whose support lies in a subspace $M$, but which take all
states in $M$ to the same final state.

\section{Teleportation and Reversible Quantum Operations}
\label{sect: teleport}

We begin this section with a brief outline of the teleportation scheme 
described in Bennett {\it et al.}\ \cite{Bennett93a}.  This scheme 
involves a sender, Alice, and a receiver, Bob.  Alice is in 
possession of two two-level systems, the input system, labeled 1, and
another system, labeled 2.  Bob is in possession of a third two-level 
system, labeled $3$. We use $|\mathord{\uparrow}\rangle$ and 
$|\mathord{\downarrow}\rangle$ to denote an orthonormal set of basis 
states for each two-level system. It should be noted that Bennett 
{\it et al.}\ extended this scheme to $n$-level systems.

Initially the composite system is prepared in a state with density 
operator $\rho \otimes \sigma$, where $\rho$ is an unknown state of 
the input system $1$, and $\sigma$ is a maximally entangled pure 
state of systems $2$ and $3$,
\begin{eqnarray}
\sigma = {1\over2}\bigl(|\mathord{\uparrow}\mathord{\downarrow}\rangle +
	 |\mathord{\downarrow}\mathord{\uparrow}\rangle\bigr)
         \bigl(\langle\mathord{\uparrow}\mathord{\downarrow}|+
         \langle\mathord{\downarrow}\mathord{\uparrow}|\bigr)\;.
\end{eqnarray}

Alice's goal is to ``teleport'' the input state $\rho$ to the target 
system, Bob's system $3$. This is done as follows. Alice performs a 
measurement on systems $1$ and $2$ in the Bell operator basis 
\cite{Braunstein92a}, which consists of four entangled states for systems 
$1$ and $2$,
\begin{eqnarray}
|\psi^{\pm}\rangle & = & {1\over\sqrt2}
   \bigl(|\mathord{\uparrow}\mathord{\downarrow}\rangle \pm
	 |\mathord{\downarrow}\mathord{\uparrow}\rangle\bigr)\;, \\
|\phi^{\pm}\rangle & = & {1\over\sqrt2}
   \bigl(|\mathord{\uparrow}\mathord{\uparrow}\rangle \pm
	 |\mathord{\downarrow}\mathord{\downarrow}\rangle\bigr)\;. 
\end{eqnarray}
Alice sends the result of this measurement, which we denote
by $i = 1, 2, 3$ or $4$, to Bob. It was shown in \cite{Bennett93a}
that there exist unitary operators $U_i$, acting only on the target
system $3$, which belongs to Bob, such that if Bob performs the 
unitary operation $U_i$ corresponding to measurement result $i$, then 
the final state of Bob's system is the input state $\rho$.  Alice has 
``teleported'' the state $\rho$ to Bob, with the help of the two bits 
of classical information necessary to tell Bob the result $i$ of her 
measurement.

We devote the remainder of this section to formulating the problem of 
teleportation generally; in particular, we show how teleportation can 
be understood in terms of reversing quantum operations.  Suppose Alice 
has possession of an {\em input system}, which we label 1, in an unknown 
input state $\tilde\rho^1$.  To avoid confusion here and throughout the 
remainder of this paper, we use a superscript to denote the appropriate 
state space for a vector or an operator; the reason for the tilde becomes 
clear shortly.  Alice might also have access to another system, which we 
label $2$.  Bob has access to the {\em target system}, which we label $3$. 
Systems $2$ and $3$ are assumed to be prepared initially in some standard 
state $\sigma^{23}$, which is assumed to be uncorrelated with 
$\tilde\rho^1$; that is, the initial state of the composite system 
consisting of 1, 2, and 3 is
\begin{eqnarray}
\tilde\rho^1\otimes\sigma^{23}\;. \end{eqnarray}
The case where Bob has access to an additional system, labeled $4$, is 
discussed briefly later in this section.

We assume that systems $1$ and $3$ are identical and thus have the same 
state space.  This means that there is a one-to-one linear map from 
the state space of 3 onto the state space of 1.  Though this map is not 
unique, we choose a particular one, thereby setting up a one-to-one 
correspondence between vectors in the state space of 3 and vectors in 
the state space of 1.  We denote this one-to-one correspondence by 
\begin{eqnarray}
|\psi^3\rangle\leftrightarrow|\tilde\psi^1\rangle\;. \end{eqnarray}
The one-to-one correspondence between vectors induces a one-to-one
correspondence between operators on 3 and operators on 1, which we
denote by $A^3\leftrightarrow\tilde A^1$. This correspondence is
given by linearly extending the map $|\psi^3\rangle \langle \phi^3|
\leftrightarrow |\tilde \psi^1\rangle \langle \tilde \phi^1|$ to all
operators on systems $3$ and $1$.  In particular, for each
state $\tilde\rho^1$ of the input system, there is a unique 
counterpart state $\rho^3$ of the target system.

The choice of a correspondence between the state spaces of 1 and 3 is 
{\it physically\/} motivated: the correspondence defines what it means 
to transport a system unchanged from the location of system 1 to the 
location of system 3.  Different procedures for performing this 
transportation lead to different correspondences.  For example, suppose
we wish to teleport the state of a spin-$1\over2$ particle from Albuquerque
to Santa Barbara.  To say what it means to teleport the state requires
a correspondence between the state spaces in Albuquerque and Santa Barbara.  
We could set up the correspondence by agreeing that the $z$ axis in each 
location lies along the local acceleration of gravity and the the $x$ axis 
along the local magnetic north or by adopting arbitrary orthogonal axes 
in the two locations.  Ordinarily we assume implicitly such a 
correspondence, as is done in the original paper on teleportation, and 
write $\tilde\rho^1=\rho^3=\rho$. 

The correspondence can be extended to a one-to-one correspondence between 
the joint state space of 2 and 3 and the joint state space of 1 and 2.  
If $|b^2\rangle|c^3\rangle$ is a product basis for the joint state space 
of 2 and 3, this one-to-one correspondence is given by
\begin{eqnarray}
|\psi^{23}\rangle=\sum_{b,c}\alpha_{bc}|b^2\rangle|c^3\rangle
\leftrightarrow
\sum_{b,c}\alpha_{bc}|\tilde c^1\rangle|b^2\rangle=
|\tilde\psi^{12}\rangle \;. \end{eqnarray}
This correspondence induces a one-to-one correspondence between operators
on the joint state space of 2 and 3 and operators on the joint state
space of 1 and 2.  

The correspondence can be extended further to a one-to-one linear map 
from the state space of the composite system 1, 2, and 3 onto itself:
\begin{eqnarray}
|\psi^{123}\rangle\leftrightarrow
|\tilde\psi^{123}\rangle=
U_{13}|\psi^{123}\rangle\;. \end{eqnarray} 
This map is accomplished by a unitary operator $U_{13}$, which acts on 
product states according to
\begin{eqnarray}
U_{13}|\tilde a^1\rangle|b^2\rangle|c^3\rangle= 
|\tilde c^1\rangle|b^2\rangle|a^3\rangle
\end{eqnarray}
and thus is called the ``swap'' operator because it swaps the states
of systems 1 and 3, while leaving system 2 alone.  The swap operator
clearly satisfies $(U_{13})^2=I^{123}$, that is, $U_{13}^\dagger=U_{13}$.
When extended to operators on the composite system, the correspondence 
becomes
\begin{eqnarray}
A^{123}\leftrightarrow\tilde A^{123}=
U_{13} A^{123} U_{13}^\dagger\;. \end{eqnarray}

Suppose now that Alice performs a measurement on systems $1$ and $2$.
This measurement is described by operators $\tilde A_{ij}^{12}\otimes I^3$, 
where the operators $\tilde A_{ij}^{12}$ are operators on the joint system 
consisting of 1 and 2, $i$ as usual labeling the result of the measurement. 
If the measurement has outcome $i$, then the unnormalized state of the 
target system $3$ after the measurement is given by 
\begin{equation} \label{cond state of 3}
\hat\rho_i^3 = \mbox{tr}_{12}
\!\left( \sum_j (\tilde A_{ij}^{12}\otimes I^3)
	        (\tilde\rho^1\otimes\sigma^{23}) 
                [(\tilde A_{ij}^{12})^{\dagger}\otimes I^3]
\right) . \end{equation}
where the caret denotes an unnormalized state.  

We now show that $\hat\rho_i^3$ is related to $\rho^3$ by a quantum 
operation, which we denote ${\cal E}_i$.  We first notice that
\begin{eqnarray} \label{sigma eqtn}
\tilde\rho^1\otimes\sigma^{23}=
U_{13}(\tilde\sigma^{12}\otimes\rho^3)U_{13}^{\dagger}\;, \end{eqnarray}
where $\tilde\sigma^{12}$ is the counterpart of $\sigma^{23}$.  
Substituting this into (\ref{cond state of 3}) gives
\begin{eqnarray} \label{eqtn: rho_i again}
\hat\rho_i^3 &=& \mbox{tr}_{12}
\Biggl( \sum_j 
                (\tilde A_{ij}^{12}\otimes I^3) \nonumber \\
&\mbox{}&\hphantom{\mbox{tr}_{12}\Biggl(}\times
                [U_{13}(\tilde\sigma^{12}\otimes\rho^3)U_{13}^{\dagger}]
                [(\tilde A_{ij}^{12})^{\dagger}\otimes I^3]
\Biggr)\;. \end{eqnarray}
The form of this equation allows us to think of $\hat\rho_i^3$ as arising
from the following process.  The composite system begins in the state
$\tilde\sigma^{12}\otimes\rho^3$, in which the joint system 1 and 2
is in the state $\tilde\sigma^{12}$ and system 3 is in the state $\rho^3$.  
After the composite system evolves under the unitary swap operator, a 
measurement is performed on the joint system 1 and 2, and then the 
joint system 1 and 2 is discarded.  This process being a measurement on 
system 3, it is not surprising that the state change from $\rho^3$ 
to $\hat\rho_i^3$ is described by a quantum operation.  We now show 
explicitly how to construct the quantum operation ${\cal E}_i$.  This 
having been done, the problem of teleportation is for Bob to reverse the 
quantum operation ${\cal E}_i$.  If the reversal can be done, then Bob 
can recover the state $\rho^3$ from the output state 
$\hat\rho_i^3={\cal E}_i(\rho^3)$ of system 3.

We write
\begin{eqnarray}
\tilde\sigma^{12}=\sum_k p_k 
|\tilde s_k^{12}\rangle \langle\tilde s_k^{12}|\;, \end{eqnarray}
where the vectors $|\tilde s_k^{12}\rangle$ make up the complete orthonormal 
set of eigenvectors of $\tilde\sigma^{12}$ in the joint space of $1$ 
and $2$. Furthermore, we let 
$\tilde\Pi_l^{12} = |\tilde P_l^{12}\rangle\langle\tilde P_l^{12}|$ be 
any complete set of orthogonal one-dimensional projectors for the joint 
system $1$ and $2$. Performing the partial trace of 
Eq.~(\ref{eqtn: rho_i again}) in the basis $|\tilde P_l^{12}\rangle$
yields
\begin{eqnarray}
\hat\rho_i^3 & = & 
\sum_{jkl} \Bigl(\sqrt{p_k}\langle\tilde P_l^{12}|
	 (\tilde A_{ij}^{12}\otimes I^3)U_{13}|\tilde s_k^{12}\rangle\Bigr )
\,\rho^3 \nonumber \\
&\mbox{}&\phantom{\sum_{jkl}}\times
	 \Bigl(\sqrt{p_k}\langle\tilde s_k^{12}|U_{13}^{\dagger}
	 [(\tilde A_{ij}^{12})^{\dagger}\otimes I^3]
         |\tilde P_l^{12}\rangle\Bigr)\;.
\end{eqnarray}
Using the single index $m$ to denote the triple $(j,k,l)$ and defining
the system~3 operators
\begin{eqnarray}
B_{im}^3 
&\equiv&\sqrt{p_k}\langle\tilde P_l^{12}|
        (\tilde A_{ij}^{12}\otimes I^3)U_{13}|\tilde s_k^{12}\rangle 
\nonumber \\
&=&\sqrt{p_k}\langle\tilde P_l^{12}|U_{13}
   (I^1\otimes A_{ij}^{23})|\tilde s_k^{12}\rangle\;,
\end{eqnarray}
we can write the output state of system~3 as
\begin{eqnarray}
\hat\rho_i^3 = \sum_m B_{im}^3 \rho^3 (B_{im}^3)^{\dagger} 
\equiv {\cal E}_i(\rho^3)\;. \end{eqnarray}
As we set out to show, $\hat\rho_i^3$ is related to $\rho^3$ by a 
quantum operation ${\cal E}_i$.

Notice that because of the sums introduced by the partial trace and the 
orthogonal decomposition of $\tilde\sigma^{12}$, the quantum operations 
${\cal E}_i$ generally are not ideal even if the measurement on 1 and 2 
is ideal.   In the next section we explore a case where the quantum
operations ${\cal E}_i$ are ideal.

For Bob to perform teleportation, he must now perform a deterministic
quantum operation ${\cal R}_i$ on system 3 such that
\begin{eqnarray}
{\cal R}_i\!\left( 
\frac{{\cal E}_i(\rho^3)}{\mbox{tr}\bigl({\cal E}_i(\rho^3)\bigr)}
\right) = 
\rho^3\;.
\end{eqnarray}
We have shown that the problem of understanding teleportation can be 
reduced to the problem of understanding how to reverse quantum operations. 
Given the work that has been done on reversing deterministic quantum 
operations that arise from decoherence, this would seem to be a useful
insight (see \cite{Schumacher96a,Shor95a,Ekert96a,Knill96a,Nielsen96a,%
Schumacher96b,Shor96a,Steane96a} for a sample of this work).

In this paper we are mainly interested in the case where Bob does the
reversal using unitary quantum operations.  Notice that if Bob had access 
to an additional system, 4, then he could perform nonunitary, but still
deterministic quantum operations in order to restore the original input
state. Using the earlier result that a necessary condition for reversing
a general quantum operation is that the operation yield no information 
about the input state, we see that a necessary condition for teleportation 
is that Alice gain from her measurement no information about the state of 
the input system.  In this paper we concentrate on unitary reversal of 
ideal quantum operations, so beyond this remark, we do not consider 
the case where Bob has access to an extra system. 

Ideal quantum operations arise naturally in the teleportation scheme
considered by Bennett {\it et al.} and might also arise in other schemes. 
In this case we seek unitary operators $U_i$ such that ${\cal E}_i$ is 
unitarily reversible by $U_i$,
\begin{eqnarray}
U_i\frac{{\cal E}_i(\rho^3)}{\mbox{tr}\bigl({\cal E}_i(\rho^3)\bigr)}
U_i^\dagger = \rho^3.
\end{eqnarray}
This is precisely the condition that is required to achieve teleportation! 

\section{Characterization of Teleportation Schemes}
\label{sect: Bennett}

In this section we consider teleportation schemes of the type introduced
by Bennett {\it et al.}\ \cite{Bennett93a}.  Suppose we have a composite 
system made up of three parts, each with the same $d$-dimensional state 
space $H$, so the state space of the composite system is 
$H^1\otimes H^2\otimes H^3$. Alice has possession of systems 1 and 2, 
and Bob has possession of system 3. In the scheme of Bennett 
{\it et al.}\ outlined earlier, $H$ is a two-dimensional state space.

The composite system is prepared in the state
\begin{eqnarray}
\tilde\rho^1 \otimes \sigma^{23}\;, \end{eqnarray}
where $\tilde\rho^1$ is any state (pure or mixed) of system $1$ and
\begin{eqnarray}
\sigma^{23}=|s^{23}\rangle\langle s^{23}| \end{eqnarray}
is a pure state of the joint system made up of systems $2$ and $3$. 
In the case considered by Bennett {\it et al.}, $\sigma^{23}$ is a 
maximally entangled pure state of systems $2$ and $3$.

Alice performs a joint measurement on systems $1$ and $2$. We assume 
that this measurement is an ideal measurement described by measurement 
operators 
\begin{eqnarray}
\sqrt{\gamma_i}\,\tilde\Pi_i^{12}\otimes I^3\;, \end{eqnarray}
where the operators 
\begin{eqnarray}
\tilde\Pi_i^{12}=|\tilde P_i^{12}\rangle \langle\tilde P_i^{12}| 
\end{eqnarray} 
are one-dimensional projectors onto the joint system made up of 1 and 2
and the $\gamma_i$ are real constants satisfying $0<\gamma_i\le1$ 
\cite{footnote}.  The measurement operators satisfy a completeness 
relation, which in this case becomes
\begin{eqnarray} \label{eqtn: Pi completeness}
\sum_i\gamma_i\tilde\Pi_i^{12}=
\sum_i\gamma_i|\tilde P_i^{12}\rangle \langle\tilde P_i^{12}| =
I^{12}\;, \end{eqnarray}
but they need not be orthogonal---that is, the vectors 
$|\tilde P_i^{12}\rangle$ can be an overcomplete set of nonorthogonal
vectors.  If the projection operators are orthogonal, then all the 
constants $\gamma_i=1$.   As remarked earlier, in the scheme of Bennett 
{\it et al.}, the operators describing the measurement are projectors 
onto an orthonormal Bell basis for the joint system $1$ and $2$.

As in the preceding section, we can write
\begin{eqnarray}
\tilde\rho^1 \otimes \sigma^{23} = 
U_{13} (\tilde\sigma^{12}\otimes\rho^3)U_{13}^{\dagger}\;, \end{eqnarray}
where $\tilde\sigma^{12}=|\tilde s^{12}\rangle\langle\tilde s^{12}|$, the
counterpart of $\sigma^{23}$, is a pure state of the joint system 1 and 2. 
The unnormalized state of the target system, given result $i$, is
\begin{eqnarray}
\hat\rho_i^3 & = & 
      \gamma_i\,\mbox{tr}_{12}\Bigl(
           (\tilde\Pi_i^{12}\otimes I^3)
           (\tilde\rho^1\otimes\sigma^{23}) 
           (\tilde\Pi_i^{12}\otimes I^3)
      \Bigr ) \nonumber \\
	     & = & 
      \gamma_i\,\mbox{tr}_{12}\Bigl(
           (\tilde\Pi_i^{12}\otimes I^3)
           U_{13}(\tilde\sigma^{12}\otimes\rho^3)U_{13}^{\dagger}
           (\tilde\Pi_i^{12}\otimes I^3)
      \Bigr ) \nonumber \\
             & = & 
      \bigl(\sqrt{\gamma_i}
      \langle\tilde P_i^{12}|U_{13}|\tilde s^{12}\rangle\bigr)
      \rho^3
      \bigl(\sqrt{\gamma_i}
      \langle\tilde s^{12}|U_{13}^\dagger|\tilde P_i^{12}\rangle\bigr)
      \nonumber \\ 
	     & = & A_i^3\rho^3 (A_i^3)^{\dagger}\;, \end{eqnarray}
where 
\begin{eqnarray}
A_i^3 \equiv 
\sqrt{\gamma_i}\langle\tilde P_i^{12}|U_{13}|\tilde s^{12}\rangle 
\end{eqnarray}
is an operator on system 3 alone.  We have shown that if the joint system
2 and 3 is initially in a pure state and if the measurement on systems
1 and 2 is an ideal measurement described by one-dimensional projectors,
then $\hat\rho_i^3$ is related to $\rho^3$ by an ideal quantum operation,
\begin{eqnarray}
\hat\rho_i^3 = {\cal E}_i(\rho^3) = A_i^3 \rho^3 (A_i^3)^{\dagger}\;. 
\end{eqnarray}

We now notice that the probability for outcome $i$ is given by
\begin{eqnarray} 
\mbox{Pr}(i) = \gamma_i\,\mbox{tr}\bigl(
           (\tilde\rho^1\otimes\sigma^{23})(\tilde\Pi_i^{12}\otimes I^3)
	   \bigr)\;, \label{eqtn: Pri}
\end{eqnarray}
which reduces to
\begin{eqnarray}
\mbox{Pr}(i) =
\mbox{tr}(\hat\rho_i^3) =
\mbox{tr}\bigl(\rho^3 (A_i^3)^\dagger A_i^3\bigr)\;.
\end{eqnarray}
{}From the result characterizing unitarily reversible ideal quantum
operations, it follows that Bob can achieve teleportation by doing
a unitary operation on system 3 if and only if $\mbox{Pr}(i)$ does not
depend on the input state $\tilde\rho^1$ of system 1.

In the original paper on teleportation \cite{Bennett93a}, it was noted
incidentally that $\mbox{Pr}(i) = \frac{1}{4}$, independent of the
input state for system 1, for each of the four possible measurement 
results.  We now see that this is in fact a sufficient condition to 
do teleportation in the scheme Bennett {\it et al.}\ were considering.  
In the original description of teleportation, the unitary operators on 
system 3 used to reconstruct the input state were given explicitly, and 
it was necessary to verify directly that these operators worked.  An 
advantage of the present approach is that much less explicit computation 
has to be done in order to verify that teleportation is possible.  Of 
course, one must construct the required unitary operators to perform 
teleportation. This is done by inverting the measurement operators 
$A_i^3$ found in the above construction.

To summarize, we have shown that the following are {\it sufficient\/} 
conditions to be able to perform teleportation:

\begin{enumerate}

\item Prepare the composite system so that the state of system 1 is 
unknown, but the state of the joint system 2 and 3 is known exactly, 
that is, is a pure state.

\item Perform a measurement on the joint system $1$ and $2$ that 
gives complete information about the {\it posterior\/} state of 
that system---that is, the joint system 1 and 2 is left in a pure 
state---but that gives {\it no\/} information about the {\it prior\/} 
state $\tilde\rho^1$ of system~1.

\end{enumerate}

\noindent
Under these circumstances, given the result $i$ of the measurement,
Bob can apply a unitary operation $U_i$ to system $3$, thereby putting
it into the state $\rho^3$, the counterpart to the initial state 
$\tilde\rho^1$ of system 1.

The teleportation scheme of Bennett {\it et al.}\ is an example of a 
scheme following this pattern. We can make some simple, yet powerful
deductions about teleportation schemes of this type using the condition that
the measurement
probabilities~(\ref{eqtn: Pri}) be independent of $\rho$.  This condition 
can be written as
\begin{eqnarray}
p_i/\gamma_i=\mbox{tr}_1 
\Bigl(\tilde\rho^1\,
        \mbox{tr}_{23}\bigl(
	(I^1\otimes\sigma^{23})(\tilde\Pi_i^{12}\otimes I^3)
	\bigr) 
\Bigr)\;, \label{eqtn: Prcond}
\end{eqnarray}
where $p_i$ is the constant value of the probability for result~$i$.  
Since Eq.~(\ref{eqtn: Prcond}) holds for all input states $\tilde\rho^1$, 
we see that
\begin{eqnarray} \label{alg condition}
(p_i/\gamma_i)I^1 &=&
\mbox{tr}_{23}\bigl((I^1\otimes\sigma^{23})
(\tilde\Pi_i^{12}\otimes I^3)\bigr) \nonumber \\
&=& \langle s^{23}|\tilde\Pi_i^{12}\otimes I^3|s^{23}\rangle \nonumber \\
&=& \langle s^{23}|\tilde P_i^{12}\rangle\langle\tilde P_i^{12}|s^{23}\rangle
\;, 
\end{eqnarray}
where it is understood that $\langle s^{23}|\tilde P_i^{12}\rangle$ acts
to the left as an operator on system 1 and to the right as an operator
on system 3.

Consider the Schmidt decomposition \cite{Peres93a} of the initial state
$|s^{23}\rangle$ of the joint system 2 and 3:
\begin{eqnarray}
|s^{23}\rangle = 
\sum_j \alpha_j |2_j\rangle |3_j\rangle\;. \end{eqnarray}
Here the vectors $|2_j\rangle$ and $|3_j\rangle$ make up orthonormal 
bases for systems 2 and 3, respectively; we choose the phases so that 
the coefficients $\alpha_j$ are real and nonnegative.  Now expand 
$|\tilde P_i^{12}\rangle$ as
\begin{eqnarray}
|\tilde P_i^{12}\rangle = 
\sum_{lm} \beta_{i,lm} |\bar1_l\rangle|2_m\rangle\;, \end{eqnarray}
where the vectors $|\bar1_l\rangle$ make up any orthonormal basis for 
system $1$.  Combining the expansions of $|s^{23}\rangle$ and 
$|\tilde P_i^{12}\rangle$ gives
\begin{eqnarray}
\langle s^{23}|\tilde P_i^{12}\rangle=
\sum_{jl}\alpha_j\beta_{i,lj}|\bar1_l\rangle\langle3_j|\;. \end{eqnarray}
Substituting this into (\ref{alg condition}), we see that the condition 
for teleportation becomes
\begin{eqnarray}
(p_i/\gamma_i)I^1 =
\sum_{jll'} 
\beta_{i,lj} \alpha_j^2 \beta_{i,l'j}^* |\bar1_l\rangle 
\langle\bar1_{l'}|\;.
\end{eqnarray}
An equivalent matrix expression is 
\begin{eqnarray} \label{eqtn: BAB}
{\bf B}_i{\bf A}^2{\bf B}_i^{\dagger} = 
(p_i/\gamma_i){\bf I}\;, \end{eqnarray}
where ${\bf A}$ is the (positive) diagonal matrix with elements 
$A_{jk}=\alpha_j\delta_{jk}$ and ${\bf B}_i$ is the matrix with elements
$B_{i,lm}=\beta_{i,lm}$.  Notice that the normalization of $|s^{23}\rangle$
can be written as $\mbox{tr}({\bf A}^2)=1$, and the normalization of 
$|\tilde P_i^{12}\rangle$ as $\mbox{tr}({\bf B}_i^\dagger{\bf B}_i)=1$.

Now write ${\bf B}_i$ in terms of a polar decomposition \cite{Peres93a},
\begin{eqnarray} 
{\bf B}_i={\bf V}_i{\bf P}_i\;, \end{eqnarray}
where ${\bf V}_i$ is a unitary matrix and 
${\bf P}_i=\sqrt{{\bf B}_i^\dagger{\bf B}_i}$ is a positive matrix.
Multiplying Eq.~(\ref{eqtn: BAB}) on the left by ${\bf V}_i^\dagger$ 
and on the right by ${\bf V}_i$, we see that
\begin{eqnarray} \label{eqtn: PAP}
{\bf P}_i{\bf A}^2{\bf P}_i = 
(p_i/\gamma_i){\bf I}\;. \end{eqnarray}
Since by assumption $p_i/\gamma_i>0$, it follows that ${\bf A}$ and 
${\bf P}_i$ both have nonzero determinants and thus are invertible.
In particular, we have that $\alpha_j > 0$ for all $j$.  Furthermore,
by manipulating Eq.~(\ref{eqtn: PAP}), we can conclude that 
\begin{eqnarray}
{\bf P}_i=\sqrt{p_i/\gamma_i}\,{\bf A}^{-1}\;. \end{eqnarray}
Then the normalization condition for $|\tilde P_i^{12}\rangle$
implies that
\begin{eqnarray}
{p_i\over\gamma_i}=
{\mbox{tr}({\bf P}_i^2)\over\mbox{tr}({\bf A}^{-2})}=
{1\over\mbox{tr}({\bf A}^{-2})}\equiv k\;, \end{eqnarray}
where $k$ is a constant independent of the measurement result $i$.
Define now, for each $i$, a new orthonormal basis for system 1 by
\begin{eqnarray}
|1_{i,j}\rangle\equiv\sum_l V_{i,lj}|\bar1_l\rangle \;, \end{eqnarray}
in terms of which the expansion of $|\tilde P_i^{12}\rangle$ becomes
a Schmidt decomposition,
\begin{equation}
|\tilde P_i^{12}\rangle=\sum_{jm}P_{i,jm}|1_{i,j}\rangle|2_m\rangle=
\sqrt k \sum_j \alpha_j^{-1}|1_{i,j}\rangle|2_j\rangle \;,
\end{equation}
with the system 2 basis in the Schmidt decomposition the same as the 
system 2 basis in the Schmidt decomposition of $|s^{23}\rangle$.

The last ingredient comes from the completeness 
relation~(\ref{eqtn: Pi completeness}): 
\begin{eqnarray}
I^1&=&\langle2_j|I^{12}|2_j\rangle \nonumber \\
&=&\sum_i\gamma_i\langle 2_j|\tilde P_i^{12}\rangle
\langle\tilde P_i^{12}|2_j\rangle \nonumber \\
&=&{1\over\alpha_j^2}\sum_ip_i|1_{i,j}\rangle\langle1_{i,j}| \;. 
\end{eqnarray}
Taking the trace of both sides gives $\alpha_j=1/\sqrt d$ for all $j$,
which implies that $k=1/d^2$ and
\begin{eqnarray}
p_i = \frac{\gamma_i}{d^2}.
\end{eqnarray}
Thus we find that 
the initial state vector of the joint system 2 and 3,
\begin{eqnarray} \label{eqtn: s23}
|s^{23}\rangle = 
{1\over\sqrt d}\sum_j |2_j\rangle |3_j\rangle\;, \end{eqnarray}
is maximally entangled, and the measurement state vectors,
\begin{eqnarray} \label{eqtn: Pi12}
|\tilde P_i^{12}\rangle &=& 
{1\over\sqrt d}\sum_j |1_{i,j}\rangle|2_j\rangle \;,
\end{eqnarray}
are maximally entangled states of the joint system 1 and 2.  

What we have shown is a complete characterization of teleportation
schemes of the type introduced in \cite{Bennett93a}.  Any maximally
entangled state of the joint system 2 and 3 can be used as the
initial state of 2 and 3, and any set of
maximally entangled states of the joint system 1 and 2 which satisfy
the completeness relation~(\ref{eqtn: Pi completeness}) can be used
to define the measurement operators on 1 and 2.  For a maximally 
entangled state, one of the orthonormal bases in the Schmidt decomposition 
can be chosen arbitrarily, so it is always possible to put all 
these maximally entangled states in the canonical form of 
Eqs.~(\ref{eqtn: s23}) and (\ref{eqtn: Pi12}), in which all the Schmidt 
decompositions share a common basis in system 2. These conditions are
both necessary and sufficient to do teleportation provided it is
assumed that the state of systems $2$ and $3$ is pure, and the POVM
elements measured by Alice are one-dimensional. It is clear that
the teleportation scheme introduced in \cite{Bennett93a} satisfies
these conditions.

\section{Conclusion}
\label{sect: conc}

We have proved a general result characterizing unitarily reversible ideal
quantum operations. In the context of quantum measurements this result 
has an intuitive physical meaning: an ideal quantum measurement is 
unitarily reversible if and only if no information about the prior 
quantum state is obtained as a result of the measurement. The 
characterization has two limitations to be addressed by further work: 
it needs to be extended to apply to any quantum operation, not just 
ideal ones, and it should allow the reversal to be performed nonunitarily, 
provided the reversal is still deterministic.

We have shown how quantum teleportation can be understood in terms
of the general problem of reversing quantum operations, thereby
demonstrating the crucial connection between teleportation and the fact 
that no information about the state to be teleported is gained during 
the process.  We have used the condition for unitarily reversing an
ideal quantum operation to characterize completely teleportation schemes
of the type introduced by Bennett {\it et al.}

\section*{Acknowledgments}

We thank H.~Barnum, C.~A.~Fuchs, and B.~Schumacher for instructive 
and enjoyable discussions about quantum information.  We thank C.~A. 
Fuchs, in particular, for suggesting to us a connection between 
reversing measurements and teleportation.  This work was supported in 
part by the Office of Naval Research (Grant No.\ N00014-93-1-0116) and 
the Phillips Laboratory (Grant No.\ F29601-95-0209).   We thank the 
Institute for Theoretical Physics for its hospitality and for the 
support of the National Science Foundation (Grant No.\ PHY94-07194).  
MN acknowledges financial support from the Australian-American Educational 
Foundation (Fulbright Commission).


\begin{thebibliography}{10}

\bibitem{Mabuchi96a}
H. Mabuchi and P. Zoller, Phys.\ Rev.\ Lett.\ {\bf 76},  3108  (1996).

\bibitem{Bennett93a}
C.~H. Bennett {\it et~al.}, Phys.\ Rev.\ Lett.\ {\bf 70},  1895  (1993).

\bibitem{Kraus83a}
K. Kraus, {\em States, Effects, and Operations} (Springer-Verlag, Berlin,
  1983).

\bibitem{Schumacher96a}
B. Schumacher, LANL e-print quant-ph/9604023, to appear in Phys.\ Rev.\ A
  (unpublished).

\bibitem{Gardiner91a}
C.~W. Gardiner, {\em Quantum Noise} (Springer-Verlag, Berlin, 1991).

\bibitem{Peres93a}
A. Peres, {\em Quantum Theory: Concepts and Methods} (Kluwer Academic,
  Dordrecht, 1993).

\bibitem{Braunstein92a}
S.~L. Braunstein, A.~Mann, and M.~Revzen, Phys.\ Rev.\ Lett.\ {\bf 68},
3259 (1992).

\bibitem{Shor95a}
P.~W. Shor, Phys.\ Rev.\ A {\bf 52},  2493  (1995).

\bibitem{Ekert96a}
A. Ekert and C. Macchiavello, LANL e-print quant-ph/9602022 (unpublished).

\bibitem{Knill96a}
E. Knill and R. Laflamme, LANL e-print quant-ph/9604034 (unpublished).

\bibitem{Nielsen96a}
M.~A. Nielsen, LANL e-print quant-ph/9606012 (unpublished).

\bibitem{Schumacher96b}
B. Schumacher and M.~A. Nielsen, to appear in Phys.\ Rev.\ A (unpublished).

\bibitem{Shor96a}
P.~W. Shor, LANL e-print quant-ph/9605011 (unpublished).

\bibitem{Steane96a}
A.~M. Steane, LANL e-print quant-ph/9605021 (unpublished).

\bibitem{footnote}
It is trivial to generalize the discussion of Sec.~\ref{sect: Bennett}
to measurement operators of the form
$$\sqrt{\gamma_i}(\tilde V_i\tilde\Pi_i^{12})\otimes I^3= 
\sqrt{\gamma_i}|\tilde Q_i^{12}\rangle\langle
\tilde P_i^{12}|\otimes I^3\;,$$
where $\tilde V_i$ is any unitary operator on the joint system 1 and 2 
and $|\tilde Q_i^{12}\rangle\equiv\tilde V_i|\tilde P_i^{12}\rangle$.
To keep the clutter of the notation at its present level, we stick to 
the simpler case in Sec.~\ref{sect: Bennett}.

\end{thebibliography}
\end{document}